\documentclass[10pt]{article}
\usepackage{csquotes}

\date{}

\usepackage{url}

\usepackage{balance}  

\usepackage{amssymb}
\usepackage{amsmath}
\usepackage{graphicx}
\usepackage[vlined,linesnumbered]{algorithm2e}
\usepackage{subcaption}

\usepackage[pdf]{pstricks}

\usepackage{amsthm}

\usepackage{color}

\usepackage{ulem}

\newcommand{\ie}{\emph{i.e.}, }
\newcommand{\eg}{\emph{e.g.}, }

\usepackage[makeroom]{cancel}
\newcommand{\nc}{$\bcancel{\mathcal{C}}$\xspace}
\newcommand{\wc}{$\mathcal{C}$\xspace}

\theoremstyle{definition}
\newtheorem{definition}{Definition}[section]

\begin{document}

\author{
  Erwan Le Merrer\\ Technicolor
  \and
  Gilles Tr\'edan\\ CNRS/LAAS
}

\title{The topological face of recommendation:\\ models and application to bias detection}

  
\maketitle

\begin{abstract}

Recommendation plays a key role in e-commerce and in the entertainment
industry. We propose to consider successive recommendations to users
under the form of \textit{graphs of recommendations}. We give models
for this representation.  Motivated by the growing interest for
algorithmic transparency, we then propose a first application for
those graphs, that is the potential detection of introduced
recommendation bias by the service provider. This application relies
on the analysis of the topology of the extracted graph for a given
user; we propose a notion of recommendation coherence with regards to
the topological proximity of recommended items (under the measure of
items' $k$-closest neighbors, reminding the "small-world" model by
Watts \& Stroggatz). We finally illustrate this approach on a model
and on Youtube crawls, targeting the prediction of "Recommended for
you" links (\ie biased or not by Youtube).



\end{abstract}

The output of recommender systems are benchmarked by researchers and
practitioners based on their precision and recall performances on test
datasets~\cite{handbook}. Yet, while those metrics have proven useful
for assessing the performances of recommenders, we find that the
\textit{graph} data-structure has not been applied for studying and
learning about the recommendations made to users (\ie the recommenders'
outputs). We argue that graph theory and the wide spectrum of graph
algorithms available for data mining complex networks can be as well
leveraged for complementing studies about recommender results.  Our
proposal is to represent the recommendations to users in either a
global \textit{graph of recommendations}, available by the service
provider at a given point in time, or as a \textit{user-graph of
  recommendations} that only captures the recommendation space to a
single user, and that can also be observed at the service or by the
user herself through the crawling of the service recommendation
interface. The extracted graph topology is thus to be leveraged for
analysis.


One application we target is related to the field of algorithmic
transparency. Some major service providers, such as Youtube, comment
on the high level implementation of their recommender, without
specifying details that would allow a transparent use by the
public~\cite{youtube}. Recently, there has been an increase in the will for
accountability of the service provided by those systems, that can be
viewed as black-boxes operating in the cloud, and that a user
interacts with by providing her profile or by calling API
operations~\cite{cscw-nous,chaintreau,stealing}.
In the setup of the observation by a user of what she gets as
recommendations, both the input (her user profile) and the output
(tens of recommendations) are very sparse in comparison to the 
dataset belonging to the service provider for analysis. In this paper,
we show as an application that observed user-graphs of
recommendations, despite their sparsity, bring interesting learning.

We first illustrate in next section the construction of a user-graph of
recommendations.

\section{Illustration: user's recommendations as a graph}

Recommendations on a website take the simple form of a set of
displayed items, for the user to interact with. We propose to go
beyond the collection of this flat item-set, by crawling from each
proposed item, recursively, up to a limited depth $h$ (for
obvious practical reasons). On the canonical example of crawling from an
item web-page (\eg video), where $k$ other related items are
recommended, and where each item is only recommended once in total, we would
obtain a balanced tree of $n_{tree}^h(k) = \frac{k^{h+1}-1}{k-1}$ nodes.

\begin{figure*}[t!]
  \vspace{-1cm}
  \begin{minipage}{0.495\linewidth}
    \hspace{-1cm}\includegraphics[angle=-100,width=1.4\textwidth]{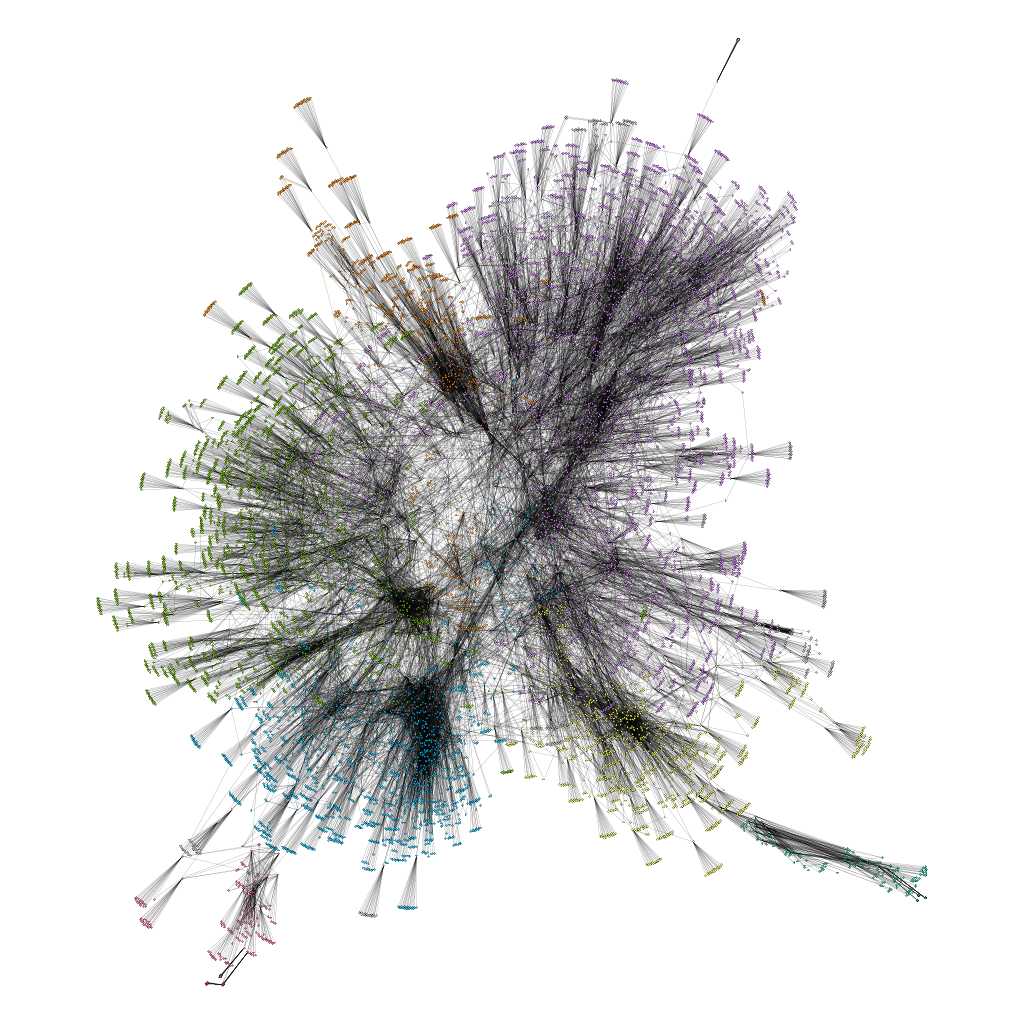}
  \end{minipage}
  \begin{minipage}{0.495\linewidth}
    \includegraphics[width=0.9\textwidth]{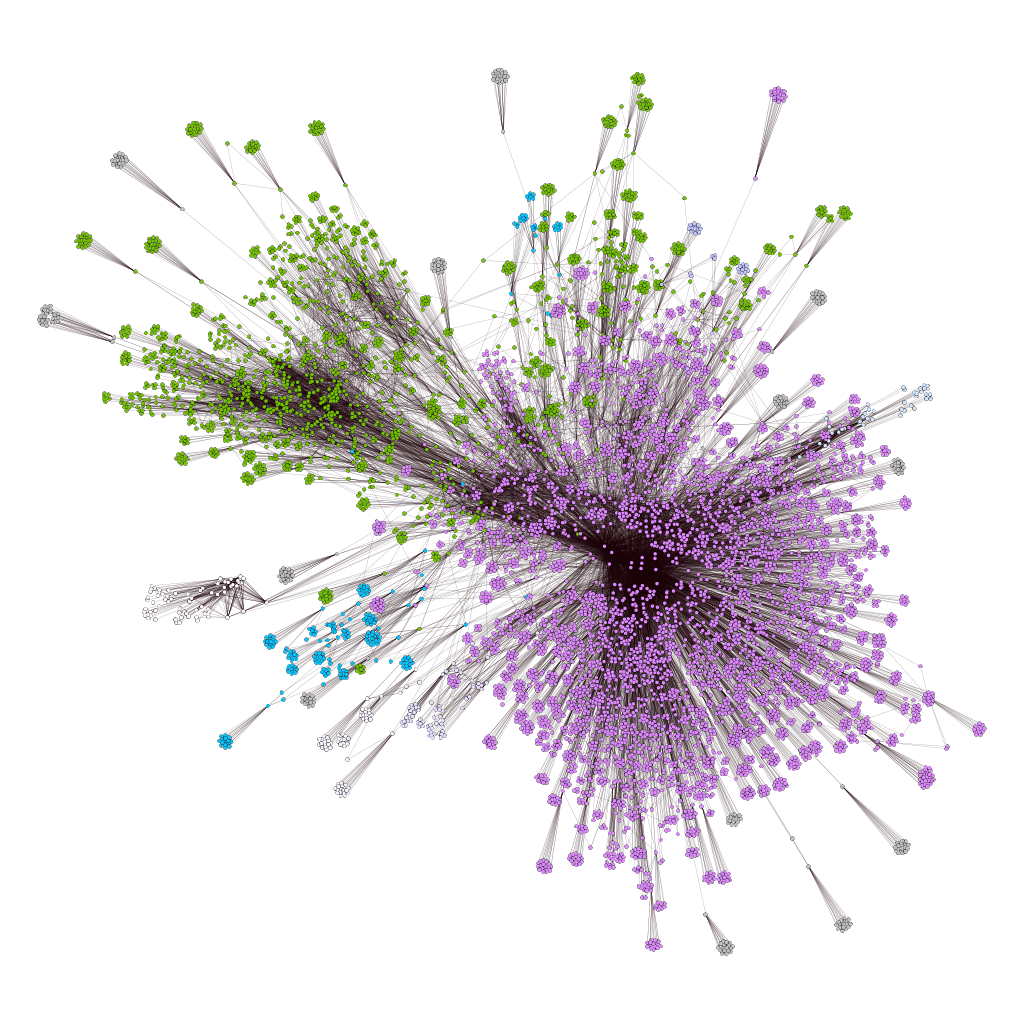}
  \end{minipage}%
        \vspace{-1.8cm}
  \caption{4-hops recommendation graphs from a Youtube video, new user (left) and returning user (right).}
  \label{fig:graphs}
\end{figure*}

We use the example of video recommendation in this paper. In the
mainstream Youtube platform, we start crawling recommended video
web-pages from a given video (a popular one from Lady Gaga), with $h=4$.
The crawling does not appear to be rate-limited, so we could collect
tens of thousands of web-pages in the order of a minute. We remark that
19 videos are recommended at each page ($k=19$). There are two clear
modes of video viewing in such a system; \textit{(i)} case \nc: a user is new to
the platform, or \textit{(ii)} \wc: she is a returning user, and thus has a history on it (recognized by Youtube
through the passing of a cookie by our web-crawler). Drastic
differences occur when building the graphs for both scenario (the two
crawls are separated by only few seconds); \nc:
$n_{\bcancel{\mathcal{C}}}^4(19) = 14,121$ nodes and has $36,435$
edges. In case \wc, $n_{\mathcal{C}}^4(19)=8,786$ nodes and has
$24,731$ edges. Both graphs are displayed on
Figure~\ref{fig:graphs}. First observation is that as
$n_{tree}^4(19)=137,561$, there is a high redundancy of
recommendations, within both \nc and \wc. Very
interestingly, crawl \wc contains around $38\%$ less nodes than \nc.
This has to be interpreted by the recommender system ``knowing'' the
user, and then trying at each video access to insist in the
recommendation of what it thinks is best suited for that user to enjoy.
In case \nc (new user), the
recommender presents videos related to the start one as well, but also
includes in its recommendations videos from different categories
(sport, news, \dots), in the probable hope to gain knowledge faster
about the user by varying its propositions.
Those phenomenons are confirmed by analyzing graph structures: a
search for main clustered components (through the modularity algorithm
with $p=5.0$) indicates $7$ components of size at least $1\%$ of the
graph size for graph from \nc, versus only $3$ for case
\wc (\ie more precise recommendations lead to fewer clusters of
interest for the returning user). Node colors corresponds to components
they belong to, on Figure~\ref{fig:graphs}. Another key difference is
the degree distribution, with $3$ nodes having more than $100$
in-neighbors for \nc, versus $10$ in the second case: 
videos assumed relevant by the system are consistently recommended to the returning user.

An immediate conclusion drawn from the structural analysis of both \nc
and \wc graphs, is that their topological comparison informs about the
degree of knowledge of the system about the observing user, even under a
limited exploration scope. Having illustrated the possibility for a
user to crawl a recommender system and analyze its output under the
form of a graph of recommendations, we now define two variants of a
general graph model for analyzing recommendations.

\section{Representation for graphs of recommendations}

The aim of recommendation is to propose suited items to users, in the
global set of available items, and based on the freshest information
available for those users. We focus for this model on item-item
recommendation~\cite{handbook}, \ie other items that are presented to a user enjoying
or reviewing a given item.


Let's imagine the service provider taking a snapshot of the
recommender's system data-structures, so that all operations on those
data-structures are frozen and observable at arbitrary time $t$. From
such a snapshot, the recommendations for a given user-item tuple
$\langle u,i \rangle$, provided by an arbitrary recommendation
algorithm, are observable or computable under the form of \eg a ranked list of
items.
\begin{definition}[Graph of recommendations]\label{gor} The directed graph $G_t=(V_t,E_t,
W_t)$ is extracted at time $t$, with $V_t$ the set of available items
(nodes), $E_t$ the set of edges (\ie the recommendations) connecting
some nodes from $V_t$, and $W_t$ the weight of those edges (\ie their
number of occurrences). Edges are thus $E_t \subseteq V_t \times V_t$,
and gathered from the system data-structures as the union of
recommendation lists for every user from user set $U_t$, at time $t$:
$E_t = \cup^{u \in U_t, i \in V_t} \langle u,i \rangle$.
\end{definition}


The recommender system's working internals, adapting to the inherent
data and user churn on the platform as the time passes by, are
triggering changes on the graph to be observed.


We now introduce the user-centric counterpart of Definition~\ref{gor}:

\begin{definition}[User-graph of recommendations]\label{ugor} The directed graph $G_t(u)=(V_t,E_t)$ is extracted at time $t$, using the same data-structures as in
  Definition~\ref{gor}, with the restriction to recommendations to a
  single given user $u$: $E_t = \cup^{u, i \in V_t} \langle u,i
  \rangle$. 
\end{definition}

$G_t$ from Definition \ref{gor} is of interest for service providers,
for instance for estimating user flows among the proposed videos (\eg
such as in the \textit{random surfer} model over web-pages linked as a
graph~\cite{surfer}), while $G_t(u)$ from Definition \ref{ugor} is of
interest for a user observation of her recommendation outputs (as
discussed in the paper sequel).\footnote{We note that these graph
  structures and their dynamics relate them to \textit{time-varying
    networks} and \textit{time-varying graphs}~\cite{Casteigts2011},
  while their introduced definitions do no fully suit the particular
  domain of recommendation (\eg no presence of a \textit{latency}
  metric $\zeta$ over graph edges).  }

\paragraph{The dynamics of graphs of recommendations}

While snapshotting data-structures to build $G_t$ or $G_t(u)$ (\ie
for a target user $u$) is expected to be relatively straightforward
for service operators with full control over their recommender system
(even in the context of distributed data-structures~\cite{snapshot}),
this is more complex for an external observer such as a user of a
platform proposing recommendations.

The observation by a user $u$ of her user-graph of recommendation may
\eg be conducted through the platform API or by a crawling the service
interface (as exposed in Section 1). Because item/user churn and
recommendation computation on a large platform is expected to occur
quickly, and because a crawl is in essence a sequential
operation, such an observation may differ from a system snapshot
$G_{t}(u)$: \textit{(i)} clearly, the worst case for an observer is
when in between the access to two items, the system has updated its recommendations, thus possibly triggering another state for the user-graph of
recommendations.
\textit{(ii)} Yet, for practical implementation
reasons of the recommender service (most notably recommendation
latency), batch-oriented pre-computations are implemented, rather than
on-demand computation of recommendations~\cite{youtube2010}; this
means that recommendations are updated for instance few times a day,
which leaves the observer with a more stable system to observe. We
assume such a practical scenario for our application based on
observations, in the sequel of this paper.

\begin{definition}[Observed user-graph of recommendations]
$\hat G_{t}^h(u,i)$ denotes a user observation of $G_{t}(u)$, where
  $t$ is indicating the arbitrary time at which the first item $i$ is
  collected to sequentially build the user-graph of recommendations,
  and $h$ is the depth of exploration away from $i$.
\end{definition}
The aim of the observer is thus to collect an observation as close as
possible from a snapshot, by \eg collecting a $\hat G_{t}^h(u,i)$ with
a small $h$ (thus referring to a local observation around the initial
item $i$, minimizing access to other items), and as quickly as
possible to mitigate recommendation re-computations. In practice, this
is the data we gather in Section 2 by using a web-crawler.


\section{An application for graphs of recommendations: detecting recommendation bias with topology}




A recently discussed topic is the influence of online medias, and
their capacity to shape opinion and user tastes based on item
recommendations. We propose a technical approach based on user-graphs
of recommendations for this question.

\paragraph{Recommender Model}

We consider recommenders that, given an item $i \in V_t$
long with some other type of information like the user profile)
return a score $s_R^i: V_t \mapsto [0,1]$ typically capturing items
similar to $i$~\cite{handbook}. The output of such a recommender
system $R$ is then exploited by a service that selects the subset of
best matching items that will get recommended when a user $u$ consults
$i$ (typically a ranking operation, as exposed in \cite{youtube}). Let $R_i(u) \subset V_t$ be
this set of recommended items to $u$ at a web-page: $\forall j\in R_i(u), j'\in
V_t\setminus R_i(u), s_R^i(u,j)\geq s_R^i(u,j')$. By selecting a
recommended item $j\in R_i(u)$, $u$ enjoys $j$, and in turn gets
recommended items similar to $j$, namely $R_j(u)$.

In this context, the user-graph of recommendations
  (Definition \ref{ugor}) for $u$ is $G_{t}(u)$, in which
  an edge $(i,j)\in E_t \Leftrightarrow j\in R_i(u)$, at time $t$.
In a system that biases recommendations,
the user is proposed certain items (for economical, or legal reasons
for instance): this translates in \textit{biased edges} toward items
from set $V_t$.


\paragraph{Bias in an observed user-graph of recommendations}

The service officially exploits the recommender $R$ (such as the one
advertised in ~\cite{youtube}). The service may add recommendation
bias in destination to user $u$ for orienting her navigation among
items; this translates into $G_t(u)$ containing biased edges. Let
$E_{t,B}$ this set of biased edges; we thus have $G_t(u)=(V_t,E_t \cup
E_{t,B})$, observed as $\hat G_{t}^h(u,i)$.


We now ask the following question: \textit{having access to $\hat
  G_{t}^h(u,i)$, can user $u$ decide whether a given edge of that
  graph is biased or not?}  Hereafter, we propose a model for
addressing this question, that is based on the small-world parallel
with graphs of recommendations.




\begin{figure*}[t!]
\centering
\begin{subfigure}[]{0.49\linewidth}
  \centering
  \includegraphics[width=1\linewidth]{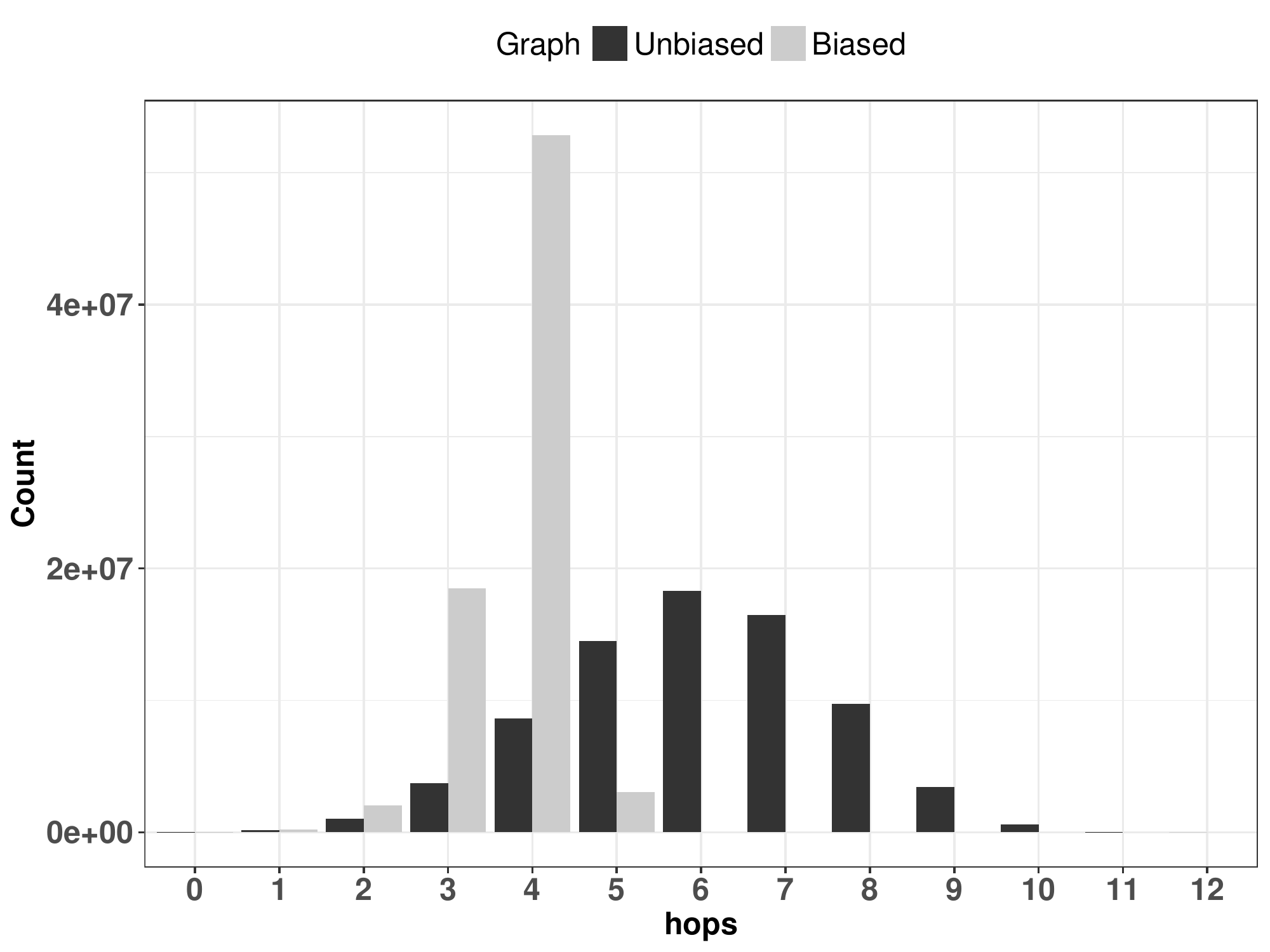}
\end{subfigure}%
\begin{subfigure}[]{.59\linewidth}
  \centering
  \hspace{-1.1cm}%
  \includegraphics[width=0.825\linewidth]{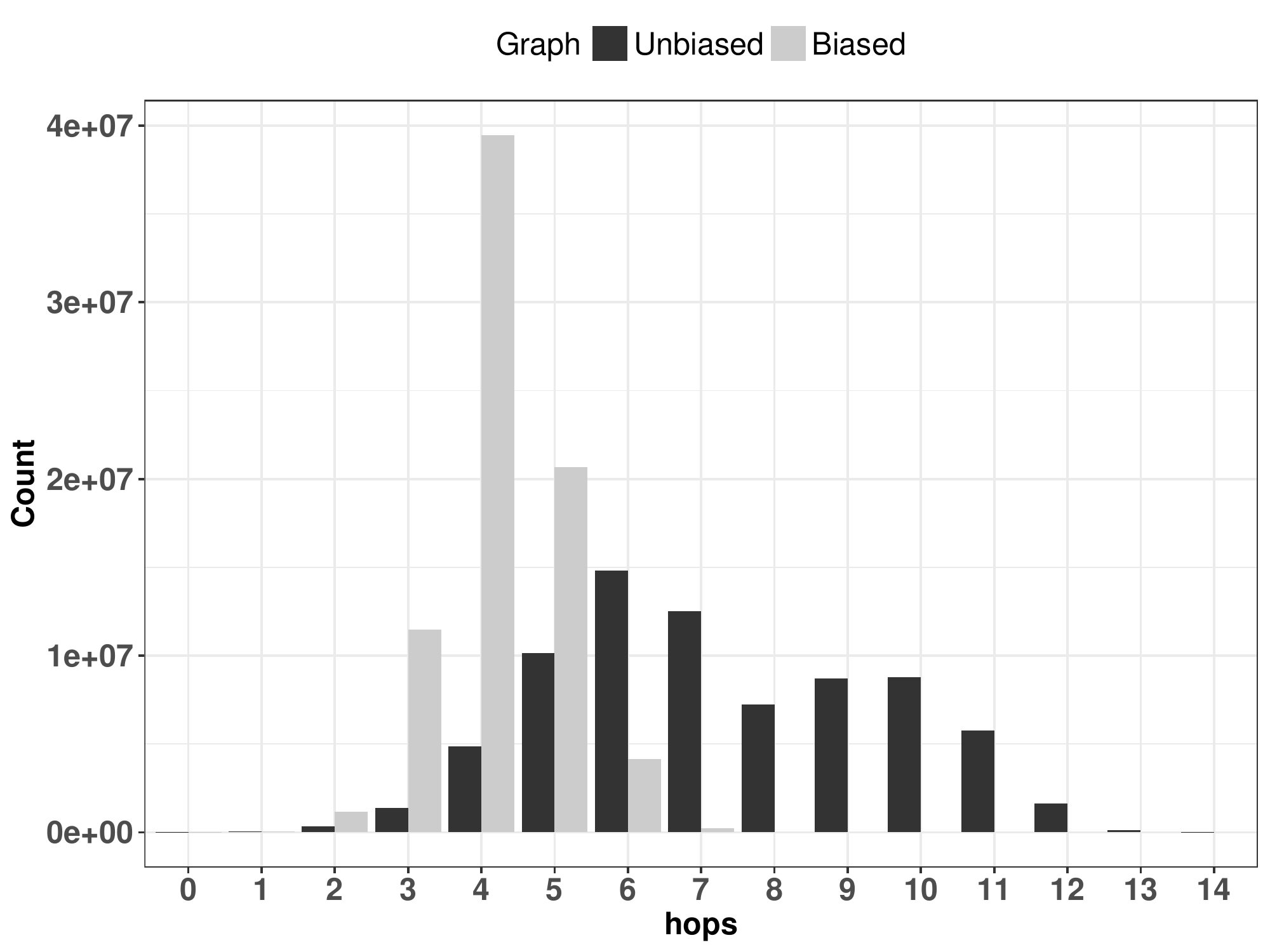}
\end{subfigure}
\caption{Distribution of graph path lengths for the experiments: model (left) and a Youtube crawl (right).}
\label{fig:hist}
\end{figure*}

\subsection{Towards algorithms for tagging long biased edges}
\label{sec:toy-model}

We propose to study potential bias in recommendation graphs, in
relation to the \textit{locality} or not of recommended items.
 For avoiding a formal - and possibly debatable - definition of bias,
 we instead state two of its most probable topological consequences on user-graph of recommendations $\hat G_{t}(u,i)$:\\
\textit{\textbf{Proposition 1} Biased edges impact graph structure}: if the service bothers to bias recommendations to a user, it is because it effectively impacts that user navigation among items. That is, users do not navigate
the same way in an unbiased graph than in a graph $G_t(u)$ containing
biased edges.  Therefore, the existence of biased edges must significantly change the
properties of the observed graph.\\
\textit{\textbf{Proposition 2} Recommenders leverage item
  proximity, and this appears in a graph observation}: most of recommenders exploit some underlying coherence
among the items. Collaborative filtering exploits correlations in
users' tastes: if users enjoying $a$ also usually enjoy $b$, $b$ will
be recommended from $a$ and vice versa~\cite{handbook}. This symmetry
translates into edge locality in $G_t(u)$, captured by \eg
clustering (as appearing through clusters on Figure~\ref{fig:graphs}). Biased edges might not rely on such property, and
therefore are less likely to produce locality. We also can argue that
if bias actually results in the proposal of usual (\ie local) items
for a user, she is not likely to consider those recommendations as biased.

In practice, machine learning algorithms associate to each item a
$d$-dimensional vector of features; a recommender then for instance rely on $k$
nearest neighbors ($k$NN) or cosine similarity on those vectors. Recommended items are
thus often close-by in the $d$-dimensional feature space, while we expect biased edges to
point to items that are relatively far in the feature space. Note that this
observation of the effect of biased edges on the topology may relate to services
providing \textit{serendipitous} recommendations~\cite{handbook} (for bringing
diversity to a user); this nevertheless arguably constitutes a form of bias
w.r.t. user habits.

\paragraph{A small-world perspective}

We now propose to identify bias using a parallel to the Watts-Strogatz
``small-world'' model~\cite{watts1998cds}. In that model, nodes have local
connectivity in a given space between them (capturing for instance a
geographical proximity), but also have so called long-range links (capturing for
instance a familial relationship, loosely related to a geographical
proximity). The consequences of these long edges are well known: they
drastically impact average path length. We argue that biased links added to the
recommender output have the same impact: provided
they are \textit{different enough} from the recommender edges, they will impact
the graph structure. To capture this degree of difference, we propose the
following model:

\begin{figure*}[t!]
\centering
\begin{subfigure}[]{0.49\linewidth}
  \centering
  \includegraphics[width=1\linewidth]{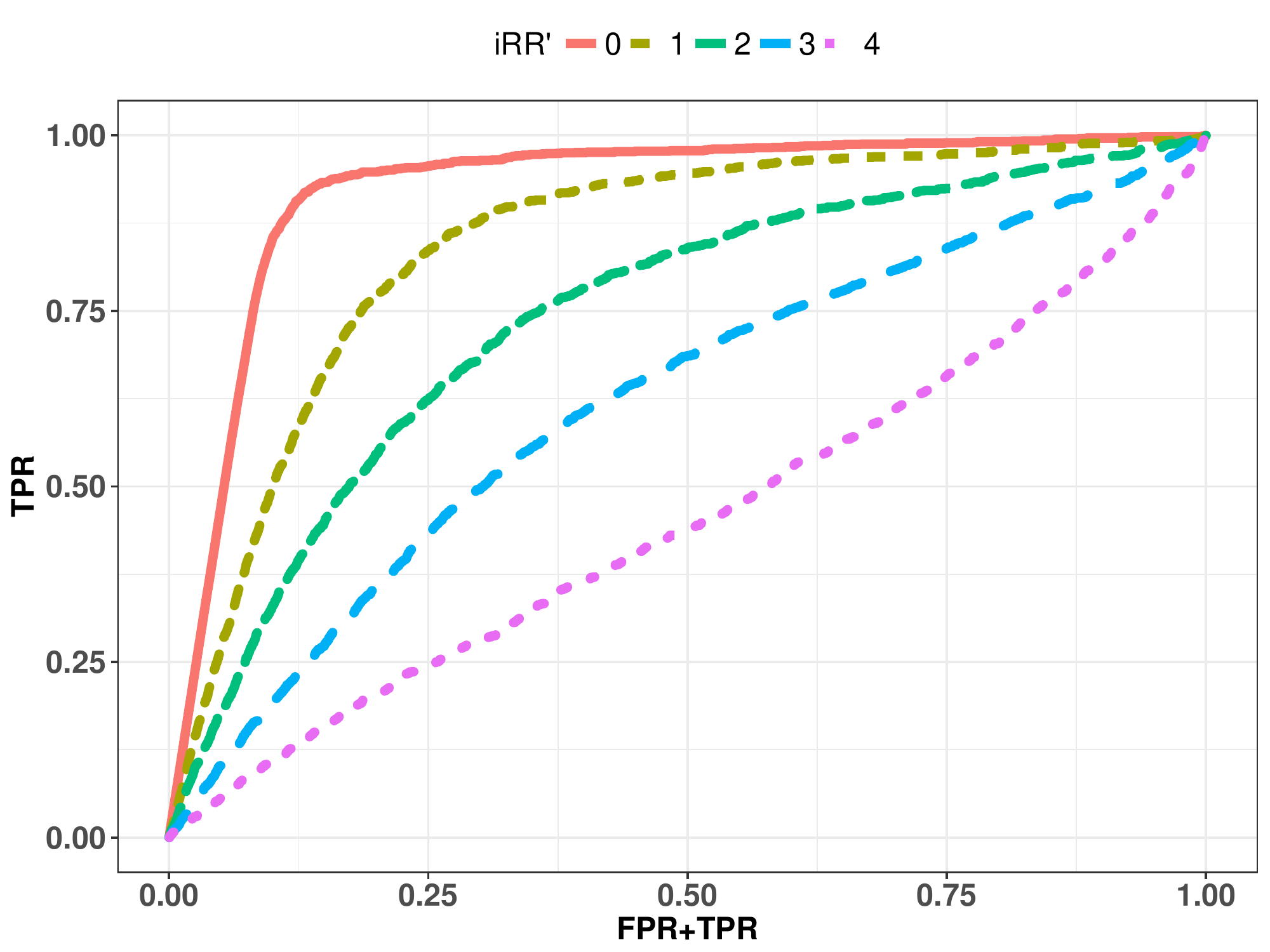}
\end{subfigure}%
\begin{subfigure}[]{.49\linewidth}
  \centering
  \includegraphics[width=1\linewidth]{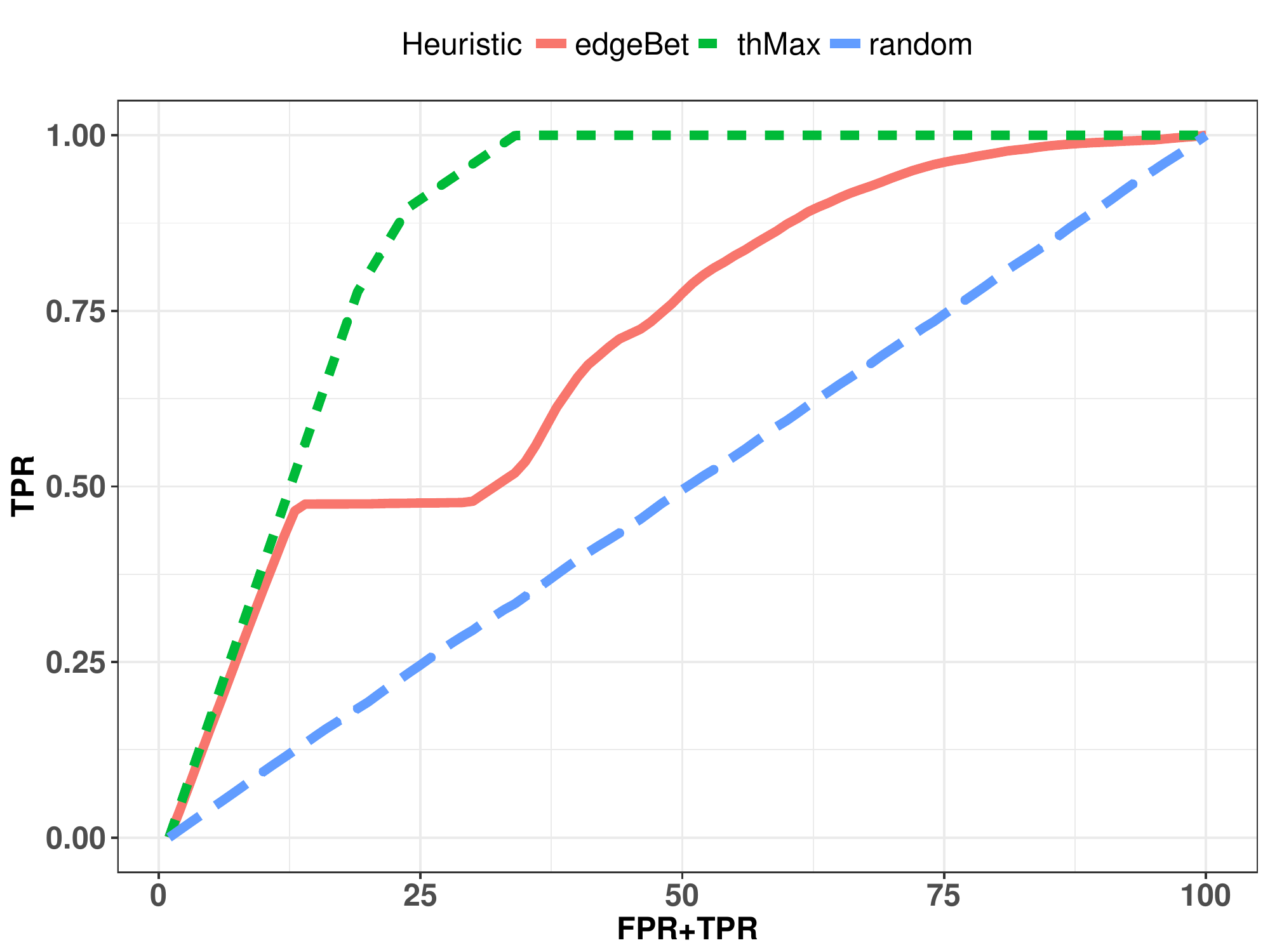}
\end{subfigure}
\caption{Feasibility of tagging biased edges: ROC curves for an edge-betweenness tagging algorithm. Model (left) and Youtube crawls (right).}
\label{fig:ROC}
\end{figure*}

\paragraph{A biased recommendation model}

Our model considers two recommenders: one \enquote{official} $R$, and one used
to issue \enquote{biased} recommendations, $R'$. To model the fact that biased
recommendations may not be completely unrelated to normal
recommendations, we use a tuning parameter $i_{RR'}$ presented
hereafter.  Let $R$ be a $k$NN recommendation system producing $k_R$
items per query ($\forall i\in V_t, |R_i(u)|=k_R$), based on the $p_i
\in \mathbb{R}^d$ feature vector: $\forall i,j\in V_t^2,
s^i_R(u,j)=||p_ip_j||_2$. Let $E_t$ be the set of produced edges.

In addition to its $p_i$ vector, each item $i\in V_t$ is associated with another
$d'$ dimensional feature vector $b_i \in \mathbb{R}^{d'}$, representing \enquote{hidden}
features (profitability, political support, \dots) leveraged to bias
recommendation. Biased edge set $E_{t,R'}$ is produced by a $k$NN recommender with
$k_B$ output using $b_i$. The user-graph of recommendation is then $G_t(u)=(V_t,E_t \cup E_{t,R'})$ in which
nodes have an out-degree of $k=k_R + k_B$.

Features are generated uniformly at random: $\forall i\in V_t:
(p_i[c])_{1\leq c\leq d} \sim U(0,1)$ and $(b_i[c])_{1\leq c\leq d'}
\sim U(0,1)$.  However, to vary the amount of bias, the dependency
between $p_i$ and $b_i$ is as following: let $0\leq i_{RR'} \leq
min(d,d')$. We set $\forall i\in V_t,\forall 1\leq c\leq i_{RR'},
b_i[c]\gets p_i[c]$. That is, if $d=d'=i_{RR'}$, both recommenders $R$
and $R'$ will produce exactly the same results (and therefore
equivalent user-graphs of recommendations). On the other hand, if
$i_{RR'}=0$, $p_i$ and $b_i$ are independent, and so are the results
of $R$ and $R'$.

\subsection{Experiments}

We use three Youtube crawls, such as the one presented as \wc (Section
1), from a returning user $u$, and from the same Lady Gaga video (with
around one month delay between each of them).  In the set of Youtube recommendations at each
page, some (around $20\%$) are tagged with the flag ``Recommended for
you'' (whereas other videos simply display their number of views). For the
experiment, we consider those recommendations as part of the biased
set we seek to tag (\ie the ground truth).  We parameter the model for
comparable topology properties: $|V_t|=8753, k_R=17,k_{B}=2$. We set
$d=d'=5$.


First, we look at the small-world parallel with introduced bias under
\textit{\textbf{Proposition 1}} on Figure~\ref{fig:hist}, where we plot the
path length distribution of respectively one $\hat G_{t}^4(u,i)$ observation (said
biased) where $i$ is the Lady Gaga video, and one $\hat G_{t}^4(u,i)
\setminus E_{t,R'}$ (said unbiased), \ie we respectively use the
YouTube full crawl, and then remove the ``Recommended for you'' edges
to obtain the unbiased graph.

We note a clear change in the graph properties when biased
edges are present: they shortcut many paths, and then cause a more
compact distribution of lengths, on both the model and the Youtube
crawl. This exact effect of long edges is a well studied feature of
small-world graphs~\cite{watts1998cds}.

Second, \textit{\textbf{Proposition 2}} is examined on Figure~\ref{fig:ROC}. For doing
so, we run a symmetric edge-betweenness centrality~\cite{Girvan11062002} algorithm on $\hat G_{t}^4(u,i)$ graphs,
for that metric is aimed at finding topologically important edges (typically linking clusters). We plot the
ROC curves, representing the probability of biased edges actually ending-up in
sorted top-result of the centrality metric (note that the the Youtube experiment, the plot is the average result over the three crawls).  For the model, independent set of edges ($iRR$=0) indicates the awaited ease to differentiate them, while a bias based on few
feature dependencies (\eg $iRR'$=1 or 2) also clearly allows for accurate edge tagging. For
Youtube, the heuristic is also significantly above a random tagging baseline, close to the maximum possible (thMax) at the beginning of the experiment (\ie top-ranked edges are indeed all biased). There is then a plateau regime (following top-ranked edges are mistagged), after which accurate tagging progresses smoothly.
For both
experiments, we conclude that tagging algorithms, to be proposed, have a clear room for providing
accurate results: if one seeks the $4491$ biased edges of the dataset, the edge-betweenness algorithm on
Youtube directly allows to identify close to $50\%$ of the biased set, at $x=12.5\%$.





\section{Related Work}

The usage of graph representations are common in the literature of
recommender systems, for analyzing their \textbf{input datasets}, and
adapt recommender algorithms accordingly. Notably, authors of
\cite{Mirza2003} propose multiple graph representations of input of a
recommender dataset (ratings, users, items): a bipartite graph of
users and items, a \enquote{social network} of users having consumed
the same items, and both the previous graphs with a so called
\textit{recommender graph}. Dataset is assumed to be the resulting of
users activity in the system following recommendations. They show the
approach useful for comparing recommender algorithms as a function of
the pairs of users they are connecting a posteriori.  Authors of
\cite{4072747} propose to build a graph of items, item categories and
users (with edges being the number of times a user consumed an item),
and perform random walks over that graph to find nodes similarities.
The temporality of user preferences (long-term and short-term) are
incorporated with approach in \cite{Xiang:2010:TRG:1835804.1835896} to build an
input (bipartite) graph that is then leveraged by a random walk-based
method to issue recommendations (similar to personalized Pagerank).
Our work differs in that we aim to analyze the \textbf{output} of
recommenders, possibly at runtime. Our models and proposed analysis
rely on a item-item graph representation.

There is a recent research focus on providing tools capable of
assessing the behavior of remote algorithms. XRay~\cite{chaintreau}
proposes a Bayesian approach for inferring which data of a user
profile, given as an input, is associated to a personalized ad to that
user.  Authors in \cite{cscw-nous} propose a graph theoretic approach
to gain understanding on which \textit{centrality} metrics are in use
by platforms that offer peer-ranking services. Work in
\cite{stealing} shows that machine learning models can be extracted by
a user from the cloud platform were they reside. Regarding recommender
systems, paper \cite{Sinha01beyondalgorithms:} exposes the users
perspective on what they wait from recommendation; interestingly,
transparency was already a concern in 2001. Yet, despite a recent rise
of concerns, they are not yet well identified tools to start addressing
the problem from an observer (user) standpoint.


\section{Conclusion}

There are two main conclusions to this study. First, graphs are an
interesting - yet vastly unexplored - tool for analyzing recommender
outputs. Graphs of recommendations may be leveraged by service
operators to compute general metrics about the consequence of their
recommendations to users, in the light of \eg Pagerank applied to the
graph of web-pages.  User-graphs of recommendations, when observed by
users themselves may also carry information: they clearly display the
item locality a service provides users with; we illustrate this
through the clustering of recommendations to a returning user and
later make the parallel with the small-world phenomenon.

Second, we believe that the representation of user-graphs of
recommendations are of important interest for the growing domain of
algorithmic transparency: if a service aims at biasing some
recommendations, the effects might be witnessed on graph topologies.

We believe these conclusions bring both analytical and algorithmic
interest in the scope of the study of personalization transparency.
Future work includes generalization to other recommender types (\eg
neural network based), and other applications of those graphs of recommendations.

\bibliographystyle{abbrv}
\bibliography{biblio.bib} 


\end{document}